\documentclass{aastex62}

\received{\today}
\submitjournal{ApJ}

\shorttitle{A Fundamental Line of Black Hole Activity}
\shortauthors{Daly et al.}
\begin{document}

\title{A Fundamental Line of Black Hole Activity}

\correspondingauthor{Ruth Daly}
\email{rdaly@psu.edu}

\author[0000-0000-0000-0000]{Ruth A. Daly}
\affil{Penn State University, Berks Campus\\
Reading, PA 19608\\}

\author{Douglas A. Stout}
\affiliation{Penn State University, Berks Campus\\
Reading, PA 19608\\}

\author{Jeremy N. Mysliwiec}
\affiliation{Penn State University, Berks Campus\\
Reading, PA 19608\\}
\nocollaboration



\begin{abstract}

Black hole systems with outflows are characterized by intrinsic physical quantities such as the outflow beam power, $L_j$, the bolometric accretion disk luminosity, $L_{bol}$, and black hole mass or Eddington luminosity, $L_{Edd}$. When these systems produce compact radio emission and X-ray emission, they can be placed on the fundamental plane (FP), an empirical relationship between compact radio luminosity, X-ray luminosity, and black hole mass. We consider a fundamental line (FL) of black hole activity written in terms of dimensionless intrinsic physical quantities: $\rm{log} (L_j/L_{Edd}) = A ~\rm{log}(L_{bol}/L_{Edd}) +B$ or equivalently $\rm{log} (L_j/L_{bol}) = (A-1) ~\rm{log}(L_{bol}/L_{Edd}) +B$, and show that the FP may be written in the form of the FL. The FL has a smaller dispersion than the FP suggesting the FP derives from the FL. Disk-dominated and jet-dominated systems have consistent best fit FL parameters suggesting they are governed by the same physics. There are sharp cutoffs at $L_{bol}/L_{Edd} \simeq 1$ and $L_j/L_{Edd} \simeq 0.2$, and no indication of a strong break as $L_{bol}/L_{Edd} \rightarrow 1$. Consistent values of $A$ are obtained for numerous samples including FRII sources, LINERS, AGNs with compact radio emission, and Galactic black holes, which indicate a weighted mean value of  $A \simeq 0.45 \pm 0.01$.  The results suggest that a common physical mechanism related to the dimensionless bolometric luminosity of the disk controls the jet power relative to the disk power. The beam power $L_j$ can be obtained by combining FP best-fit parameters and compact radio luminosity for sources that fall on the FP.

\end{abstract}



\section{Introduction} \label{sec:intro}

The fundamental plane (FP) of black hole activity  
was introduced and studied extensively 
by Merloni, Heinz, \& Di Matteo (2003) and 
Falcke, K\"ording, \& Markoff (2004). Subsequently, 
the FP has been studied with additional sources by many  
groups including K\"ording, Falcke, \& Corbel (2006), G\"ultekin et al. 
(2009), Bonchi et al. (2013), Saikia, K{\"o}rding, \& {Falcke} 
(2015), and Nisbet \& Best 
(2016), hereafter NB16. The FP encapsulates the 
relationship between compact radio luminosity, 
X-ray luminosity, and black hole mass for black hole systems 
and provides a good description of the data over a very large range
of black hole mass.  As has been
discussed by many authors, such as 
Heinz \& Sunyaev (2003), Merloni et al. (2003), Falcke et al. (2004), 
K\"ording, Fender, \& Migliari (2006), Merloni \& Heinz (2007), 
Fender (2010), de Gasperin et al. (2011), and Saikia, K\"ording, \& 
Falcke (2015), to name a few, 
the radio luminosity, $L_R$, obtained by combining the radio flux density
and the radio frequency, is likely to be related to the beam power 
or luminosity in directed kinetic energy powering the outflow, $L_j$.  
And, the X-ray luminosity, $L_X$, is likely to be related to the 
bolometric luminosity, $L_{bol}$  of the accretion disk. 
The relationships frequently discussed in the literature including
the papers noted above are $L_j \propto L_R^{\alpha}$, where the beam power of 
the outflow is defined as the energy per unit time, $L_j \equiv dE/dt$,  
ejected by the source
in the form of kinetic energy, and  
$L_{bol} \propto L_{X} (2-10 \hbox{ keV})$.  
In addition, Saikia et al. (2015) introduce the 
bolometric FP that replaces $L_X$ with $L_{bol}$ 
where $L_{bol}$ is obtained from the OIII luminosity of sources.  
This allows sources with accretion disk luminosities determined with 
tracers other than X-ray luminosity to be placed on the fundamental plane. 

It is important to understand the relationship between the intrinsic
physical variables that describe black hole systems to 
determine whether they are all governed by the same physics or whether 
there are a variety of mechanisms that produce collimated outflows. As 
discussed above, given that many different types of sources with outflows 
fall on the FP and that the FP covers such a broad range of black hole mass, 
it is expected that the underlying physics governing the properties of the 
sources is similar. However, the precise relationship between the 
underlying intrinsic source parameters and empirically determined FP 
parameters has not
been fully characterized. Here, we suggest an ansatz that allows a
determination of the  
connection between the underlying intrinsic source quantities 
($L_j$, $L_{bol}$, and $L_{Edd}$), and 
the quantities used to define the FP ($L_R$, $L_X$, and black hole mass). 
This also allows the relationships between  
intrinsic physical variables to be obtained and studied, and 
compared with model preditions, which are typically cast in terms of
intrinsic source quantities. This ansatz is presented in section 2. 

In addition, many black hole systems have empirical estimates of the bolometric 
accretion disk luminosity obtained with tracers other than 
X-ray emission or beam power estimates obtained
with tracers other than compact radio emission, and thus can not be
placed directly on the FP. 
For these sources, it is important to compare 
their properties with those of sources on the FP 
to determine whether they are 
governed by similar or different physical mechanisms. 
Such a comparison has important
implications for the physics of the outflows and the accretion disk properties 
relative to the black hole masses for different types of sources. 
To determine whether the relationship between outflow properties, 
accretion disk properties,
and black hole mass or Eddington luminosity 
are consistent with those indicated by 
the FP, it is helpful to see if
the FP may be rewritten in terms of other quantities
such as Eddington normalized 
beam power and Eddington normalized bolometric luminosity. 
The fact that the FP is valid over a very large black hole mass range
suggests that it can be written in terms of these dimensionless 
luminosities. With this representation, more sources can be compared, 
and the range of values
of dimensionless beam power and dimensionless bolometric luminosity and
the relationship between these quantities can be studied. 
In section 2, it is shown that the FP is likely the empirical manifestation
of an underlying relationship between the dimensionless beam power and
dimensionless bolometric accretion disk luminosity.  
Applications to data sets are 
presented in section 3. The results and implications are 
discussed in section 4, and summarized in section 5.  

\section{Analysis}

The FP may be written in the form of eq. (1), where
the units and precise definitions of $L_R$, $L_X$, and $M$ vary from paper 
to paper, and a, b, and c are empirically determined constants.
Using the different definitions of $L_R$, $L_X$, and the black hole mass 
$M$ in each paper, 
it is possible to convert these definitions to a common scale, which was done
by NB16. Here, we follow the definitions presented by NB16:  
\begin{equation}
\rm{log} ~L_R = a~\rm{log}~L_{X,42}  + b~\rm{log}~ M_8 + c~,
\end{equation}
where $L_R$ is the 1.4 GHz radio power of the compact radio source in 
$\hbox{erg s}^{-1}$, $L_{X,42}$ is the (2 - 10) keV X-ray luminosity of 
the source in units of 
$10^{42} \hbox{ erg s}^{-1}$, and $M_8$ is the black hole mass in units
of $10^8 M_{\odot}$; conversions to and from other wavebands are discussed by NB16. 
NB16 present results obtained with a sample 
of 576 LINERS and summarized the results of Merloni et al. (2003),
K\"ording et al. (2006), G\"ultekin et al. (2009), Bonchi 
et al. (2013), and Saikia et al. (2015).

Sources with a huge range of black hole mass, about nine orders of magnitude, 
follow the functional form of eq. (1), suggesting that 
the relationship between the intrinsic physical quantities  
$L_j$, $L_{bol}$, and black hole mass $M$ (or $L_{Edd}$) scale 
in such a way that the overall relationship is maintained. 
If the equation that describes the relationship   
between intrinsic physical
variables is written in terms of dimensionless quantities
then it is scale-invariant and 
we expect the relationship between intrinsic 
variables to remain valid for all scales.  
The simplest equation written in terms 
of dimensionless quantities with a form similar
to that of the fundamental plane is 
\begin{equation}
\rm{log}\left({L_j \over L_{Edd}}\right) = A ~\rm{log} 
\left({L_{bol} \over L_{Edd}}\right) +B,
\end{equation}
where $L_{Edd}$ is the Eddington luminosity, 
$L_{Edd} \simeq 1.3 \times 10^{46} M_8 ~\hbox{erg s}^{-1}$, and 
$A$ and $B$ are constants. So, the goal is to determine whether
eq. (1) can be written in the dimensionless 
and Eddington-normalized form of eq. (2). 

As described in section I, 
it is convenient to write the bolometric luminosity of the accretion 
disk as 
\begin{equation}
L_{bol} = \kappa_{X} ~ L_{X} (2-10 \hbox{ keV})
\end{equation}
where $\kappa_{X}$ is a constant, and the beam power $L_j$ as 
\begin{equation}
\rm{log} L_j = C ~\rm{log} ~L_R  + D ~,
\end{equation}
where $C$ and $D$ are constants, and
$L_{bol}$, $L_X (2-10 \hbox{ keV})$, $L_j$,  
$L_R$, and $L_{Edd}$ have units of $\hbox{erg s}^{-1}$.  
Substituting eqs. (3) and (4) into eq. (1) indicates that
\begin{equation}
\rm{log}\left({L_j \over L_{Edd}}\right) = aC~\rm{log}L_{bol} -(1-bC)~\rm{log}L_{Edd} + \kappa
\end{equation}
where $\kappa = D + C(c -  42~ a - 46.11~ b - a~\rm{log} \kappa_X)$. 

Eq. (5) may be written in the form of eq. (2) when $aC = (1-bC)$; that is, when 
$C = (a+b)^{-1}$. Thus, if it turns out that $C = (a+b)^{-1}$ then the FP, 
given by 
eq. (1), may be rewritten in the mass-scale-invariant form given by eq. (2). 
In this case, it is easy to show that $A = a/(a+b)$ and $B = \kappa$. 

Three approaches are used to determine whether the value of $C$ is equal to $(a+b)^{-1}$. 
First, $C$ is computed using $C = (a+b)^{-1}$, and the values obtained for
different data sets are shown to be consistent. The values
are compared with those indicated by independent  
theoretical studies and independent 
studies of radio sources, and it is shown that there is good agreement
between the empirical values, $C = (a+b)^{-1}$, theoretically 
predicted values, and values based on independent studies of radio sources. 
Second, the values of $aC$ and $(1-bC)$ are computed
using the value of $C$ indicated by independent theoretical studies and independent 
studies of radio sources, and the
difference between $aC$ and $(1-bC)$ is shown to be consistent with zero. 
And, third, again using the value of $C$ indicated by 
independent theoretical studies and
independent studies of radio sources, 
the ratio of $aC$ to $(1-bC)$ is computed and is shown to 
be consistent with unity. These results indicate that 
$aC$ is equal to $(1-bC)$, thus $C$ does equal $(a+b)^{-1}$, 
implying that eq. (1) may be written in the form of eq. (2). 
Details of the three analyses just described are 
presented in section 2.1 and are summarized in section 2.2.  

\begin{table*}
\begin{minipage}{140mm}
\scriptsize
\caption{Study of the Parameters Discussed in Section 2}   % title of Table
\label{tab:comp}        % is used to refer this table in the text
%\centering                          % used for centering table
\begin{tabular}{lllllllll}   % centered columns (4 columns)
\hline\hline                    % inserts double horizontal lines
(1)&(2)&(3)&(4)&(5)&(6)&(7)&(8)&(9)\\
Sample & $C={1 \over (a+b)}$& $aC$ & $(1-bC)$ & $aC-(1-bC)$&${aC \over (1-bC)}$ & $A = {a \over (a+b)}$ & $B = \kappa$& $D$ \\
\hline
{NB16(1)%
\footnote{All input values are from Table 3 of NB16, with the first line here
corresponding to the first line of that Table. Other values follow in the same 
order at presented in Table 3 of NB16 beginning with Merloni et al. (2003) 
[M03]; K\"ording et al. (2006) [K06]; G\"ultekin et al. (2009) [G09]; 
Bonchi et al. (2013) [B13]; and Saikia et al. (2015) [S15].}}
&$ 0.75 \pm 0.07 $&$ 0.46 \pm 0.05 $&$ 0.51 \pm 0.07 $&$ -0.05 \pm 0.09 $&$ 0.90 \pm 0.18 $&$ 0.49 \pm 0.05 $&$ -1.87 \pm 2.09 $&$ 14.93 \pm 2.61 $ \\
M03&$ 0.72 \pm 0.08 $&$ 0.43 \pm 0.08 $&$ 0.45 \pm 0.07 $&$ -0.02 \pm 0.11 $&$ 0.95 \pm 0.25 $&$ 0.43 \pm 0.05 $&$ -2.80 \pm 1.99 $&$ 15.86 \pm 3 $ \\
K06 &$ 0.72 \pm 0.26 $&$ 0.45 \pm 0.28 $&$ 0.47 \pm 0.21 $&$ -0.02 \pm 0.36 $&$ 0.96 \pm 0.75 $&$ 0.46 \pm 0.19 $&$ -2.70 \pm 2.03 $&$ 15.76 \pm 10.1 $ \\
K06 &$ 0.74 \pm 0.11 $&$ 0.40 \pm 0.05 $&$ 0.45 \pm 0.13 $&$ -0.04 \pm 0.14 $&$ 0.91 \pm 0.30 $&$ 0.42 \pm 0.06 $&$ -2.52 \pm 1.96 $&$ 15.58 \pm 4.03 $ \\
G09 &$ 0.69 \pm 0.14 $&$ 0.48 \pm 0.09 $&$ 0.45 \pm 0.19 $&$ 0.03 \pm 0.21 $&$ 1.07 \pm 0.51 $&$ 0.46 \pm 0.10 $&$ -3.90 \pm 2.04 $&$ 16.96 \pm 5.43 $ \\
G09 &$ 0.69 \pm 0.06 $&$ 0.44 \pm 0.07 $&$ 0.42 \pm 0.06 $&$ 0.02 \pm 0.09 $&$ 1.05 \pm 0.25 $&$ 0.43 \pm 0.05 $&$ -3.88 \pm 1.98 $&$ 16.94 \pm 2.38 $ \\
B13 &$ 0.93 \pm na $&$ 0.28 \pm 0.04 $&$ 0.52 \pm na $&$ -0.24 \pm na $&$ 0.54 \pm na $&$ 0.36 \pm na $&$ 5.07 \pm 1.86 $&$ 7.99 \pm na $ \\
S15 &$ 0.80 \pm 0.29 $&$ 0.45 \pm 0.28 $&$ 0.57 \pm 0.14 $&$ -0.11 \pm 0.32 $&$ 0.80 \pm 0.54 $&$ 0.51 \pm 0.18 $&$ na \pm na $&$ na \pm na $ \\
S15 &$ 0.61 \pm 0.13 $&$ 0.59 \pm 0.21 $&$ 0.42 \pm 0.14 $&$ 0.17 \pm 0.26 $&$ 1.41 \pm 0.71 $&$ 0.50 \pm 0.11 $&$ -7.58 \pm 2.11 $&$ 20.64 \pm 5 $ \\
\hline 
\end{tabular}
\end{minipage}
\end{table*}

\subsection{Details of the Analysis}

The values of $C = 1/(a+b)$ are listed in Table 1 for the values of 
$a$ and $b$ summarized in Table 3 of NB16. 
The nine values of $C$ obtained are all consistent; 
they range from about 
0.61 to 0.93 with a median value of 0.72 and unweighted mean value of $0.74$;
all of the sources are within one sigma of these mean and median values, 
as expected based on the consistency of the values of $a$ and $b$ 
listed by NB16. As discussed below,
the mean and median values of $C$ are consistent with 
theoretical predictions and independent studies of radio 
sources. 

Theoretical studies such as those by Heinz \& Sunyaev (2003) and 
Merloni \& Heinz (2007) 
indicate that the value of $C$ is expected to be $12/17 \simeq 0.71$ when 
the radio spectral index of the compact radio source is close to zero 
and the radio emission is not affected by Doppler beaming and boosting 
due to bulk motion. 
This agrees with the value of $0.71 \pm 0.03$ indicated by the 
independent studies of radio sources, discussed below.
The fact that the values of $C$ indicated by the best fit values of
$a$ and $b$ (see column 2 of Table 1) are consistent with that 
predicted theoretically and indicated
by independent studies of radio sources suggests that
relativistic beaming is likely to have a small effect on the compact radio
emission for most of the sources used to construct the fundamental plane.
   
An independent estimate of $C$, and an independent estimate of $D$, 
may be obtained by considering independent 
studies of radio sources. Using the empirically determined 
relation between core radio 
luminosity and extended radio luminosity for radio galaxies presented 
by Yuan \& Wang (2012) and applying the conversion of extended 
radio luminosity to beam power presented by Willott et al. (1999), we obtain 
$C = 0.71 \pm 0.03$ and $D = 14.2 \pm 1.0$. Note that this is only used to 
confirm in general terms that $C$ is consistent with $(a+b)^{-1} \simeq 0.7$; 
the results of Willott et al. (1999) are not used to compute the beam power 
for any of the sources presented in this paper. 

To further test whether eq. (5) can be written in the form of
eq. (2), the values of $aC$ and $(1-bC)$ and their difference and ratio
are computed and listed in columns (3), (4), (5), and (6) of 
Table 1, where the value of $C$ indicated above of $0.71 \pm 0.03$ is used 
to compute values for columns (3) through (6).
The difference listed in column (5) 
is consistent with zero for all of the samples. 
And, the ratio $aC/(1-bC)$ listed in column (6) is consistent with unity. 
Thus, eq. (1) can be written in the form of eq. (2). 

\subsection{Summary of the Analysis}
These considerations indicate that 
eq. (1) may be written in the form of eq. (2), 
where $A= a/(a+b)$, $C=1/(a+b)$, and $B = \kappa$. 
In terms of understanding the physics of the sources and 
constraining models that describe the sources, 
$A$ is the most important parameter since
eq. (2) implies that 
$L_j \propto L_{bol}^A ~ M^{(1-A)}$ $\propto L_{bol}^A ~ L_{Edd}^{(1-A)}$. 
The parameter 
$A$ may be obtained using best fit parameters  $a$ and $b$, or
by fitting to the fundamental line (FL) given by eq. (2). 
The value of $A$ obtained from the best fit values of $a$ and $b$ 
are listed in column (7) of Table 1 for nine samples of sources, and 
all values are consistent within the uncertainties.  Values of $A$ obtained 
from eq. (2) by fitting directly to Eddington normalized 
luminosities are discussed in section 3.

The y-intercept of the FL is described by $B$. 
The value of $B$ may be obtained from $B = \kappa$, which   
requires values of $C$, $D$, $\rm{log} \kappa_X$ as well as
best fit values of $a$, $b$, and $c$. To compute $B$, 
the value of $C$ listed in column (2) of Table 1 is used, as is 
$D=14.2 \pm 1.0$ obtained as described above. 
To obtain an estimate
for $\rm{log}\kappa_X$, the bolometric luminosity of each source
with a measured (2-10 keV) X-ray luminosity 
listed by Merloni et al. (2003) was obtained using eq. (21) from Marconi et al.
(2004); this indicates a mean value of $\rm{log}\kappa_X = 1.18 \pm 0.03$, or
$\kappa_X \simeq 15.1$, 
which is in good agreement with the value obtained by Ho (2009). 
It is easy to show that propagating
the uncertainties of all parameters, the uncertainty of $c$ has an
insignificant impact on the uncertainty of $B$, so this 
term is not included in computing the uncertainty of $B$. 
With these values, $B$ is obtained and listed in column (8) of Table 1.

In addition, we may write $D = B -C(c-42~ a -46.11~b -a ~\rm{log}\kappa_X)$
so that a value of $D$ may be obtained for each sample if an estimate
of the value of $B$ is available, given the value 
$\rm{log}\kappa_X$ described above and the value of $C$ listed
in column (2) of Table 1. Daly (2016) studied a sample of 97
FRII AGN for which beam powers, bolometric
luminosities, and Eddington luminosities were available and fitting to 
eq. (6) obtained
a value of $B = -1.14 \pm 0.06$. Values of $D$ obtained with this value
of $B$ are listed in column (9) of Table 1. It is easy to show that propagating
the uncertainties of all parameters, the uncertainty of $c$ has an
insignificant impact on the uncertainty of $D$, so this 
term is not included in computing the uncertainty of $D$. 

\begin{table*}
\begin{minipage}{140mm}
\scriptsize
\caption{Best-fit Parameters to Eq. (2), as Described in Section 3}   % title of Table
\label{tab:comp}        % is used to refer this table in the text
%\centering                          % used for centering table
\begin{tabular}{llllll}   % centered columns (4 columns)
\hline\hline                    % inserts double horizontal lines
(1)&(2)&(3)&(4)&(5)&(6)\\
Sample & N & A & B & {Dispersion of FL
\footnote{Obtained assuming all of the dispersion is due to 
$\rm{log}(L_j/L_{Edd})$ and decreases by the factor $1/\sqrt{2}$ if the 
error is equally weighted between this parameter and 
$\rm{log}(L_{bol}/L_{Edd})$. The uncertainties of the best
fit parameters have been adjusted to bring the reduced chi-squared to one. }}
&Dispersion of FP \\
\hline
NB16&576 LINERS &$0.43 \pm 0.02$&$-1.08 \pm 0.08$&0.47&0.73\\
M03&80 AGN + 36 GBH& $0.45 \pm 0.03$&$-0.94 \pm 0.10$& 0.65&0.88\\
M03&80 AGN & $0.41 \pm 0.04$&$-1.34 \pm 0.14$& 0.71\\
{D16
\footnote{From Daly (2016).}}
&97 FRII AGN & $0.44 \pm 0.05$&$-1.14 \pm 0.06$&0.38\\
S15&102 GBH &$0.47 \pm 0.02$&$-1.37 \pm 0.04$&0.19\\
S15&102 GBH ($\kappa_X = 5$)&$0.47 \pm 0.02$&$-1.15 \pm 0.04$&0.19\\
Combined Sample&855 sources
\footnote{Includes 576 LINERS, 80 AGN, 97 FRII AGN, and 102 GBH obtained with $\kappa_x = 5$ for GBH and $\kappa_x = 15.1$ for all other sources as described in the text.}
&$0.41 \pm 0.01$&$-1.19 \pm 0.04$& 0.47\\
Combined Sample&855 sources
\footnote{Includes 576 LINERS, 80 AGN, 97 FRII AGN, and 102 GBH obtained with the same value of $\kappa_x = 15.1$ for all sources as described in the text.}
&$0.39 \pm 0.01$&$-1.27 \pm 0.04$& 0.48\\
\hline
\end{tabular}
\end{minipage}
\end{table*}

\begin{figure}
    \centering
    \includegraphics[width=80mm]{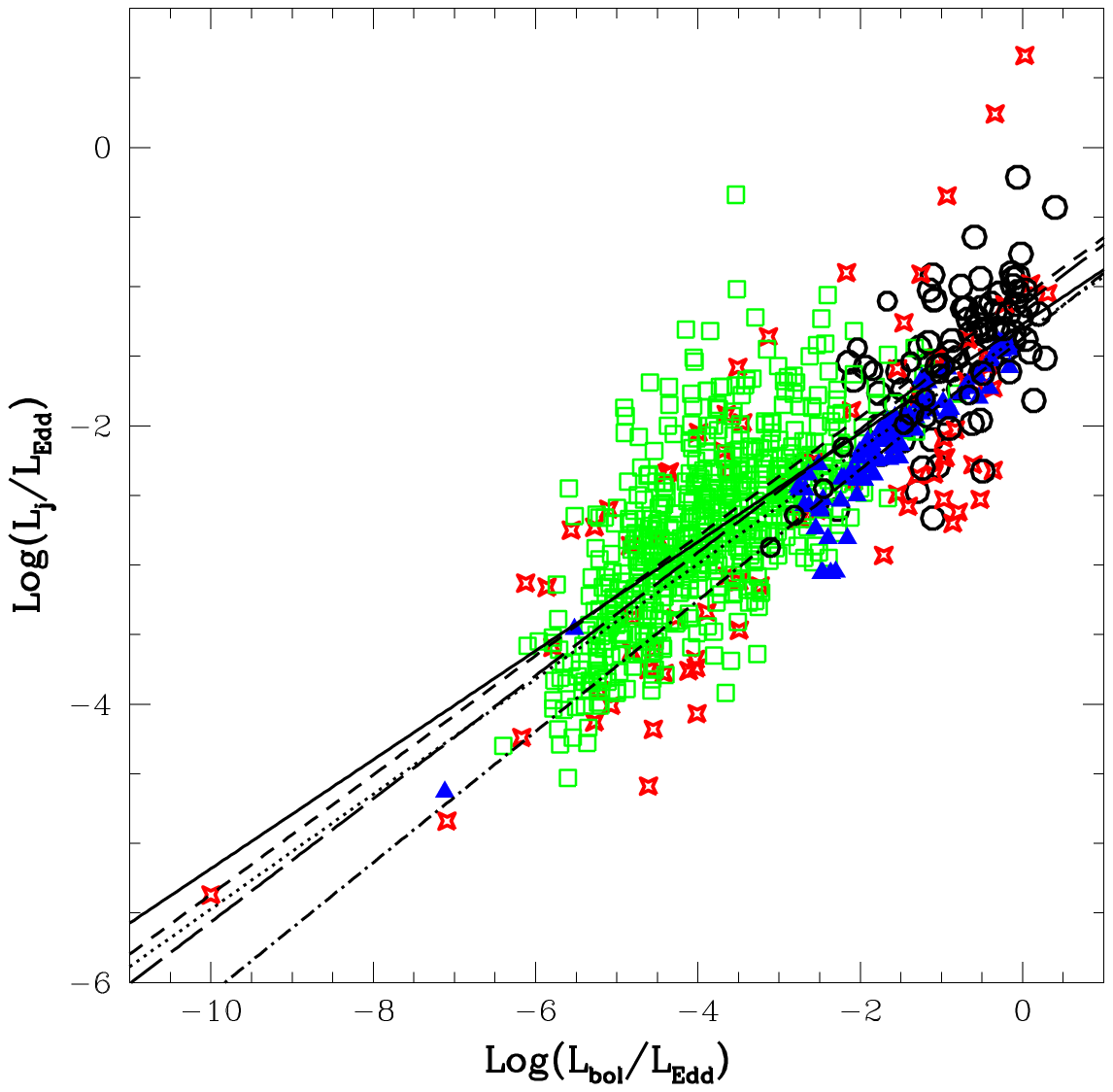}
\caption{The log of the Eddington normalized beam power  
is shown versus the log of the Eddington normalized 
bolometric accretion disk luminosity 
for 576 LINERS (open green squares), 80 AGN (open red stars), 
102 GBH measurements (solid blue triangles), and 97 powerful extended radio 
sources (open black circles); values of quantities are obtained as described
in Section 3. The best fit lines are:  medium dashed line (576 LINERS);  
dotted line (80 AGN); dot-dash line (102 GBH); 
long-dashed line (97 FRII AGN); and solid line (all 855 measurements).  
These symbols are used throughout the paper. 
Best fit parameters to eq. (2) for the quantities shown here 
are listed in Table 2 and discussed in Section 3. }
		  \label{fig:F1}
    \end{figure}

\begin{figure}
    \centering
    \includegraphics[width=80mm]{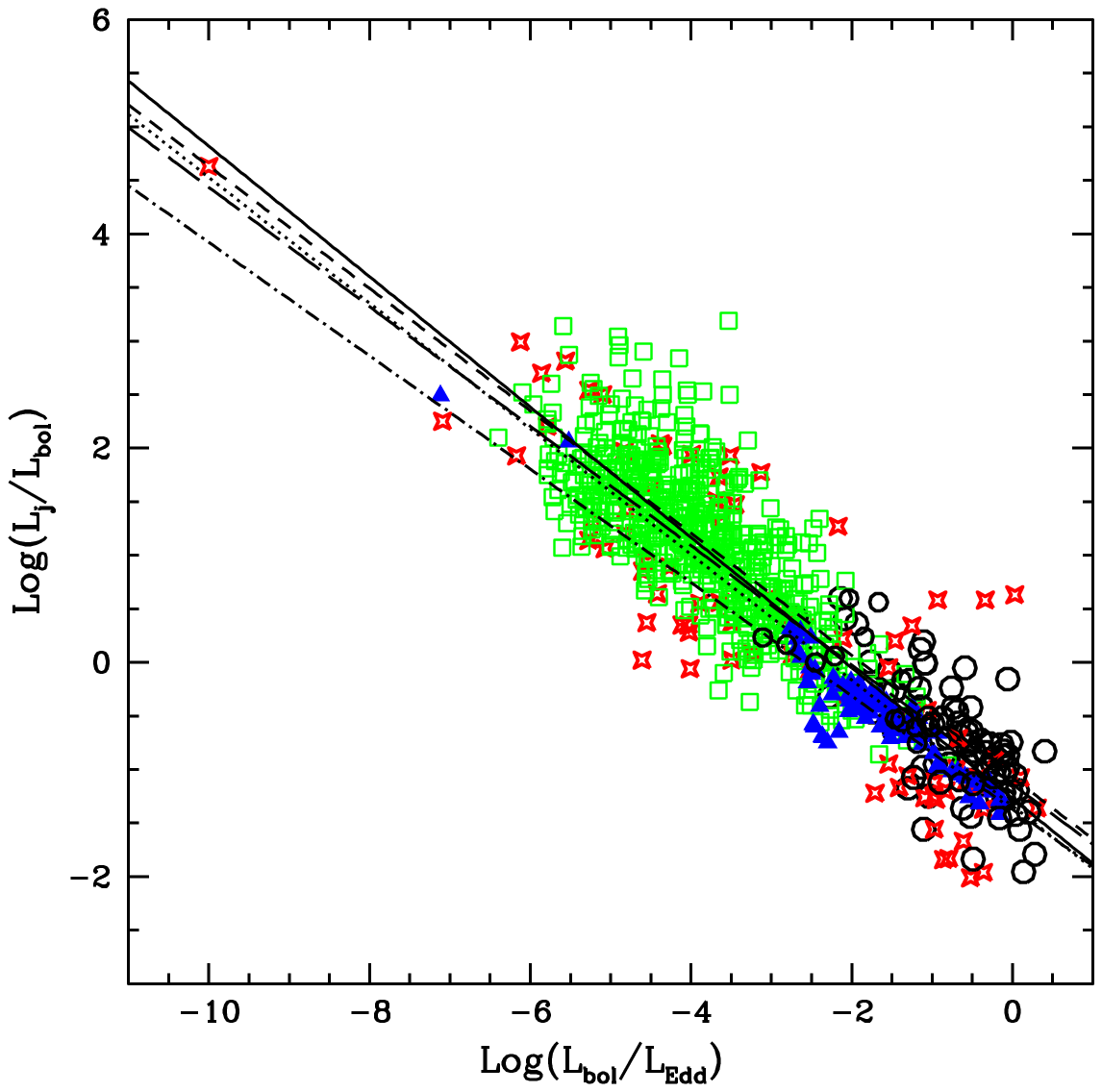}	 
\caption{The log of the ratio of beam power to bolometric 
accretion disk luminosity 
is shown versus the log of the Eddington normalized bolometric 
luminosity for the sources shown in Fig. 1. See eq. (6) and the discussion 
in section 4. }
		  \label{fig:F2}
    \end{figure} 

\section{Applications} 

The FL, given by eq. (2) or (6), may be studied when values of 
$L_j$, $L_{bol}$, and $L_{Edd}$ are available for sources. There
are numerous well-studied methods to estimate the latter two, but
there are only a few methods to estimate $L_j$. The results presented in section
2 indicate that the beam power $L_j$ may be obtained using eq. (4)
for sources that lie on the FP and have best-fit FP parameters a, b, 
and c determined, since both $C$ and $D$ in eq. (4) depend upon the best-fit FP 
for that type of source. That is, each type of source may have different
best fit FP parameters and hence different values of $C$ and $D$; 
once the best fit FP parameters are determined for that type of source,  
eq. (4) may be used to solve for the beam power of individual sources of
that type. For example, given the source types compiled and studied 
by NB16, the value of $C$ is listed in column 2, and that of $D$ is listed
in column 9 of Table 1. In addition, there are many types of sources that have
beam power determinations that do not have compact radio emission and
hence can not be placed on the FP. For example, O'Dea et al. (2009) 
used multi-frequency radio images to determine the beam power of 
powerful classical double (FRII) radio sources by applying the equations of 
strong shock physics and showed that this method allows a determination 
of the beam power that is very insensitive to assumptions. 
Notably, the beam power obtained does not depend upon assumptions
related to the distribution of energy between relativistic particles
and magnetic fields since the offsets from minimum energy conditions cancel 
out when the beam power is written in terms of empirically determined
quantities. Results obtained for FRII sources for which values of 
$Lj$ were obtained using this method will be discussed below, and the
results of fits to eqs. (2) and (6) are described in detail by Daly (2016).  

To obtain and study $L_j/L_{Edd}$ and $L_{bol}/L_{Edd}$ for sources with 
best fit FP parameters, 
samples were selected from the list of NB16 in such a way as to 
avoid duplicate measurements of any given source while
maintaining the integrity of each sample. That is, 
some sources are common to different samples, and rather than
removing sources from samples, representative samples were 
selected such that the samples do not have observations in common. 
The samples considered include the 576 LINERS 
presented by NB16, the 80 AGNs, which are all compact radio sources, 
presented by Merloni et al. (2003), and the
102 measurements of Galactic black holes (GBH) that are X-ray binaries 
presented by Saikia et al. (2015). 
The value of $L_{bol}$ is obtained for each source 
using eq. (3) with the value of $\rm{log}~\kappa_X$ described 
above (one additional value is considered for the GBH), 
black hole mass is converted to $L_{Edd}$ 
with the relation listed above, and $L_j$ is obtained using eq. (4) 
given the values of 
$C$ and $D$ listed in columns (2) and (9) of Table 1. A value of
$D$ is not available for the Saikia et al. (2015) fit that included
GBH (see row 8 of Table 1), but the fit of NB16(1) has 
values of $a$ and $b$ that are consistent with those of Saikia et al. 
(2015) and this fit goes right through the GBH points, 
so the NB16(1) values of $C$ and $D$ are applied to that sample; in 
addition, a value of $\kappa_X = 5$ suggested by 
Saikia et al. (2015) is also considered for the GBH.  
All values of $L_R$ are scaled to 1.4 GHz using the 
conversion factors of NB16. In addition, 
97 FRII AGN from Daly (2016) are included;
independent values of 
$L_j$, $L_{bol}$, and $L_{Edd}$ are available for these sources. These 
samples allow a comparison between sources of various types.  

Results obtained with these four samples and the combined 
sample are shown in Fig. 1 and are summarized in Table 2; the 
results shown in Fig. 1 are obtained for 
$\rm{log}\kappa_X = 1.18 \pm 0.03$ (i.e. $\kappa_X \simeq 15.1$) 
for all sources, and the values of $L_{bol}$ decrease
by about a factor of 3 for the GBH for $\kappa_X \simeq 5$. 
All fits are unweighted.  
To estimate the dispersion of $y = \rm{log}(L_j/L_{Edd})$, $\sigma_y$, we assume
that all of the dispersion in the fit is due to the uncertainty in $y$, and
following Merloni et al. (2003) and G\"ultekin et al. (2009), 
we assume the uncertainty is the same
for each point. Setting the reduced $\chi^2$ to unity then provides an 
estimate of $\sigma_y$. If there are additional sources of error, such as 
errors in $x = \rm{log} (L_{bol}/L_{Edd})$, then $\sigma_y$ will decrease since
the total dispersion $\sigma_T^2 = \sigma_y^2 + \sigma_x^2$. For example,
if $\sigma_x = \sigma_y$, then the 
dispersion will decrease by a factor of $1/\sqrt{2}$ from the values
listed in column (5) of Table 2. 

The values of the dispersion of the FL 
listed in column (5) of Table 2 can be compared with 
values of the dispersion of the FP, which are available for two  
of the samples studied, and these are listed in column (6) of Table 2. 
The fundamental line (FL) described by eq. (2) 
has a smaller dispersion than the FP. 
This is another indication that eq. (2) provides a good 
description of the data. 
Indeed, it suggests that the FL, which is closely tied to  
intrinsic source properties, leads to the FP, which is closely
tied to empirical source properties. That is, it suggests that the 
FP derives from or is the result of the FL. 

The results presented in Table 2 for the combined sample
support the idea (Saikia et al. 2015) that the conversion factor
$\kappa_X$ is about three times larger for AGN than it is for GBH
since the dispersion of the combined sample is a bit smaller when 
the value of $\kappa_X$ is larger for AGN than it is for GBH. 

Fig. 1 indicates that 
eq. (2) provides a good description of the data over about 10 orders
of magnitude in $L_{bol}/L_{Edd}$, from about $10^{-10}$ to about $1$, and over
about 5 orders of magnitude in $L_j/L_{Edd}$, from about $10^{-5.5}$ to 
about $0.3$.  Interestingly, the relationship seems to remain valid 
all the way up to values of $L_{bol}/L_{Edd} \rightarrow 1$.
Within uncertainties, 
the relationship between the dimensionless parameters indicated by the FL
is the same 
for all types of sources studied including LINERS, AGNs that are compact
radio sources, GBH, and FRII AGNs.  In addition, 
the values of $A$ indicated by best-fit fundamental plane parameters 
a and b, listed in column (7) of Table 1, are consistent with those 
obtained by a direct fit to the data, listed in Table 2. 

\section{Discussion}

The FL, given by equation (2), may be re-written as 
\begin{equation}
\rm{log}\left({L_j \over L_{bol}}\right) = (A-1) ~\rm{log} 
\left({L_{bol} \over L_{Edd}}\right) +B.
\end{equation}
This may be interpreted as the efficiency of the beam power
relative to accretion disk power as a function of Eddington normalized
disk luminosity. It is interesting to 
consider the range of values of this parameter, and results
obtained with the samples discussed above are shown
in Fig. 2. Here $y = \rm{log}(L_j/L_{bol})$,  and the dispersion of
this parameter for each sample and for the combined sample 
is the same as those quoted in Table 2 for $y = \rm{log}(L_j/L_{Edd})$;  
best fit values of the slope $(A-1)$, shown in Fig. 2, 
can easily be obtained from the values of $A$ listed in 
Table 2. 

Values of $L_j/L_{bol}$ extend over about 7 orders 
of magnitude from about $10^{-2} - 10^{5}$,  
and the relationship between $L_j/L_{bol}$ and $L_{bol}/L_{Edd}$ is similar
for all classes of objects studied. 

The transition from sources with
$L_j > L_{bol}$, considered to be ``jet dominated,'' 
to those with $L_j < L_{bol}$, considered to be ``disk dominated,''
occurs at about
$L_{bol}/L_{Edd} \sim 10^{-2}$, which agrees with the results obtained
by Best \& Heckman (2012) and NB16. This is consistent
with previous results that find that ``jet dominated'' 
sources have low values of $L_{bol}/L_{Edd}$, while those that 
are ``disk dominated'' have high values of this parameter
(e.g. see the summaries of Fender 2010; Best \& Heckman (2012); 
Heckman \& Best 2014; Yuan \& Narayan 2014; and NB16).  

There is no change or break in the relationship between the 
ratio $L_j/L_{bol}$ and $L_{bol}/L_{Edd}$ at $L_j/L_{bol} \sim 1$; that 
is sources with $L_j/L_{bol} < 1$ and those with  $L_j/L_{bol} > 1$ 
have the same relationship between $L_j/L_{bol}$ and $L_{bol}/L_{Edd}$.
This indicates that there is little or no difference between the 
physics of ``disk dominated'' and ``jet dominated'' sources for the types of
radio loud sources studied here. These results suggest that they 
are governed by the same physics and are described by the same equations. 

As noted above, there appears to be a rather sharp cut-off of sources at 
$L_{bol}/L_{Edd} \sim 1$ indicating that sources with outflows that fall
on the FP and FL do not have super-Eddington luminosities. In addition, there
does not seem to be a strong change in the relationship between 
$L_j/L_{bol}$ and $L_{bol}/L_{Edd}$ 
as $L_{bol}/L_{Edd} \rightarrow 1$.  

The relationship obtained between $L_j/L_{bol}$ and $L_{bol}/L_{Edd}$ 
for the FRII AGN studied by Daly (2016) is the same as that obtained
here for many other types of sources. Daly (2016) showed that this relationship
is consistent with models in which the outflow is powered by the spin of the 
black hole, such as the Blandford-Znajek mechanism (Blandford \& Znajek 1977). 
The same reasoning applies here, which implies that 
outflows from the other types of systems studied here,
including LINERS, AGNs that are compact radio sources, and GBH have outflow and 
disk properties that are also consistent
with models in which the outflow is powered by the spin of the black hole. 
That is, the value of $A$ obtained here is consistent with
predictions in models with  
$L_{bol} \propto \epsilon ~ \dot{M} \propto \epsilon ~\dot{m} ~M$, 
and $L_j \propto \dot{m} ~M~f(j)$, where 
$\dot{M}$ is the mass accretion rate, $\dot{m} \equiv \dot{M}/\dot{M}_{Edd}$
is the dimensionless mass accretion rate, 
$\dot{M}_{Edd} \equiv L_{Edd} c^{-2}$ is the Eddington accretion rate,  
$\epsilon$ is a dimensionless efficiency factor, and $f(j)$ is a function
of black hole spin (see Daly 2016).   
These equations imply that for a value of $A$ of 0.5,
$\epsilon \propto \dot{m}$, and for a value of $A$ of 0.45, 
$\epsilon \propto \dot{m}^{1.2}$ (see Daly 2016 for 
a detailed discussion).  
The equation for $L_j$ is that expected in the Blandford-Znajek (1977) 
model of spin energy extraction if the magnetic field $B$ is given 
by $B^2 \propto (\dot{m}/M)$, and this is expected to be the case
for several types of accretion disks, such as ADAF and MAD accretion 
disks (e.g.  Tchekhovskoy et al. 2010; Yuan \& Narayan 2014). Note that 
these results are consistent with those
obtained by Fender, Gallo, \& Jonker (2003), who find $L_j \propto L_X^{0.5}$.

Even though the maximum value of $L_{bol}/L_{Edd}$ is close to one, 
the maximum value of $L_j/L_{Edd}$ is about (0.1 - 0.3) (see Fig. 1). 
Still, this is a large value for $L_j/L_{Edd}$, indicating very 
powerful outflows relative to the Eddington luminosity. The fact 
that this is less than one and in the range of 0.1 to 0.3 
may be related to the limit that, at
most, 29 \% of the mass-energy of a black 
hole can be due to the spin energy of the hole, 
so at most 29 \% of the mass-energy can 
be extracted electromagnetically or with any other 
mechanism (e.g. Blandford 1990).

\section{Summary}
The fact that the FP is 
manifestly scale invariant suggests that it 
represents a relationship between the Eddington normalized beam
power and Eddington normalized bolometric luminosity of the black
hole system. In section 2 it is shown that the 
dimensionless representation given by eq. (2) is 
consistent with independent emipirical and theoretical studies of radio sources
(see columns 2, 5, and 6
of Table 1). Given that eq. (1) may be written in the form of eq. (2),
the value of $A$ may be written in terms of the best fit FP
parameter values $a$ and $b$: $A = a/(a+b)$, and these values
are listed in Table 1. Another way to determine
the value of $A$ is described in section 3: the beam power $L_j$ of a source
may be determined using eq. (4) given the values of $C$ and $D$ listed
in Table 1 for sources of that type, and the value of $L_{bol}$ may 
be obtained using eq. (3). A direct fit to eq. (2) is then possible since
it is easy to convert black hole mass to $L_{Edd}$. The values of $A$ obtained
by a direct fit to the data are presented in Table 2. 

In terms of understanding the 
physics of the sources and constraining models that describe the
sources, $A$ is of paramount importance since eq. (2) implies that
$L_j \propto L_{bol}^A~M^{(1-A)}$. All nine samples studied are consistent
with a single value of $A$ and yield a mean value of
$0.45 \pm 0.02$ (see column (7) of Table 1); this is  
consistent with the value of $0.44 \pm 0.05$ obtained by Daly (2016) 
for FRII AGN. The mean value of $A$ for the four samples 
that make up the ``combined sample'' listed in Table 2 is 
$0.45 \pm 0.01$. However, the fit to the combined sample indicates
a smaller value of $A$, which is likely due to the different y-intercepts
of the samples causing offsets that modify the value of $A$ for the fit to 
the combined sample.  

Two equivalent forms of the FL, given by eqs. (2) and (6),
are studied directly using four data sets that include LINERS, AGNs that
are compact radio sources, GBH, and FRII AGN. 
All categories of source have consistent best fit values
of $A$ and $B$ (see Figs. 1 and 2 and Table 2), and the FL has a smaller
dispersion than the FP, supporting the longstanding interpretation 
of the FP as arising from a relationship between the intrinsic 
physical variables that describe black hole systems. 
This representation allows a study of the
range of values of $L_j/L_{Edd}$, $L_{bol}/L_{Edd}$, and 
$L_j/L_{bol}$, and the relationships between these parameters.
For the sources studied $L_j/L_{Edd}$ ranges from about 
$10^{-5.5}$ to about 0.3;  $L_{bol}/L_{Edd}$ ranges from about 
$10^{-10}$ to about 1; and $L_j/L_{bol}$ ranges from about 
$10^{-2}$ to about $10^5$. 
These results indicate that the 
sources transition from jet dominated to disk dominated, which 
is defined to occur at $L_j/L_{bol} \sim 1$, at a value of 
$L_{bol}/L_{Edd} \sim 10^{-2}$, which is consistent with the value
obtained by others (e.g., Best \& Heckman 2012; NB16).  
Interestingly, the relationships between the dimensionless parameters 
$L_j/L_{bol}$ or $L_j/L_{Edd}$ and $L_{bol}/L_{Edd}$ seem
to remain valid over the full range of values of  $L_{bol}/L_{Edd}$
studied, including $L_{bol}/L_{Edd} \sim 1$, and over the full range
of values from $L_j/L_{bol} << 1$ to $L_j/L_{bol} >> 1$. 
The values of $A$ and $B$, particularly $A$, describe the physics
of the sources and have implications
for models of black hole systems. The fact that one
dimensionless equation, eq. (2) or eq. (6),  
describes all sources, including sources
that do not have compact radio emission, over such large
ranges of dimensionless parameters, suggests that a common
physical mechanism set primarily by $L_{bol}/L_{Edd}$ 
regulates the beam power relative to the
accretion disk power in these systems. Note that this includes the 
possibility that $L_{bol}/L_{Edd}$ may be impacted by $L_j/L_{Edd}$ through
feedback mechanisms. 

\section*{Acknowledgments}

Thanks are extended to Philip Best, David Nisbet, and the 
referee for very helpful comments and suggestions. Daly would like
to thank the Aspen Center for Physics for hosting the March
2016 meeting, the 2016 summer workshop on black hole
physics, and the January 2018 meeting where this work was discussed; 
in particular, thanks
are extended to Norm Murray, Syd Meshkov, Pepi Fabbiano,
Martin Elvis, Christine Jones, Bill Forman, Rosie Wyse,
Rachel Webster, and Garth Illingworth for helpful conversations.
This work was was supported in part by Penn 
State University and performed in part at the Aspen 
Center for Physics which is support by National Science Foundation
grant PHY-1066293.  Undergraduate students Douglas Stout and Jeremy 
Mysliwiec would like to acknowledge the use of facilities at Penn State Berks 
during the summer of 2016 when they were involved with this work.



\end{document}